\documentclass[authoryear,final,5p,twocolumn]{elsarticle}

\usepackage{graphicx}
\usepackage[utf8]{inputenc}

\usepackage{amssymb}
\usepackage{amsmath}
\usepackage{wasysym}

\usepackage{color}

\journal{Icarus}

\newcommand{\highlight}[1]{#1}

\begin{document}

\begin{frontmatter}



\title{A critical analysis of shock models for chondrule formation}


\author[label1,label2]{Sebastian M. Stammler \corref{cor1}}
\ead{stammler@uni-heidelberg.de}
\author[label1]{Cornelis P. Dullemond}

\cortext[cor1]{Corresponding author}

\address[label1]{Heidelberg University, Center for Astronomy, Institute of Theoretical Astrophysics, Albert-Ueberle-Straße 2, 69120 Heidelberg, Germany}
\address[label2]{Member of the International Max Planck Research School for Astronomy and Cosmic Physics at the Heidelberg University}

\begin{abstract}
In recent years many models of chondrule formation have been proposed. One of
those models is the processing of dust in shock waves in protoplanetary disks.
In this model, the dust and the chondrule precursors are overrun by shock waves,
which heat them up by frictional heating and thermal exchange with the gas.

In this paper we reanalyze the nebular shock model of chondrule formation
and focus on the downstream boundary condition. We show that for large-scale
plane-parallel chondrule-melting shocks the postshock equilibrium
temperature is too high to avoid volatile loss. Even if we include radiative
cooling in lateral directions out of the disk plane into our model (thereby
breaking strict plane-parallel geometry) we find that for a realistic
vertical extent of the solar nebula disk the temperature decline is not fast
enough. On the other hand, if we assume that the shock is entirely optically
thin so that particles can radiate freely, the cooling rates are too high to
produce the observed chondrules textures. Global nebular shocks are therefore
problematic as the primary sources of chondrules.
\end{abstract}

\begin{keyword}
Disks \sep Meteorites \sep Radiative transfer \sep Solar Nebula \sep Thermal histories

\end{keyword}

\end{frontmatter}

% \linenumbers

\section{Introduction}
\label{sectionIntroduction}

The origin of chondrules is one of the biggest mysteries in meteoritics.
These 0.1$\cdots$1 mm sized silicate once molten droplets, abundantly found in chondritic
meteorites, must have cooled and
solidified within a matter hours \citep[e.g.][]{hewins05}. From short-lived radionuclide chronology data
\citep[e.g.][]{villeneuve09} it is known that this must
have taken place during the first few million years after the start of the
solar system, during the phase when the sun was still likely surrounded by a
gaseous disk (the ``solar nebula''). What makes chondrule formation
mysterious is that this few-hour cooling time is orders of magnitude shorter
than the typical few-million-year time scale of evolution of the solar
nebula. Chondrules can thus not be a natural product of the gradual
cooling-down of the nebula. Instead, chondrules must have formed during
``flash heating events'' of some kind -- but which kind is not yet known. 
There exists a multitude of theories as to what these flash heating events
could have been. \citet{boss96} and \citet{ciesla05} give nice overviews of
these theories and their pros and cons. So far none of these theories has
been universally accepted.

One of the most popular theories is that nebular shock waves can melt small
dust aggregates in the nebula, causing them
to become melt droplets and allowing them to cool again and solidify
\citep{hood91}. The origin of such shocks could, for instance, be
gravitational instabilities in the disk \citep[e.g.][]{boley08} or the effect
of a gas giant planet \citep[e.g.][]{kley12}. Detailed 1-D models of the
structure of such radiative shocks, and the formation of chondrules in them,
were presented by \citet{iida01}, \citet{desch02}, \citet{ciesla02} or
\citet{morris10}. These models show that such a shock, in an optically thick
solar nebula, would lead to a temperature spike in the gas and the dust that
lasts for only a few seconds to minutes with cooling rates of up to
$10^3$ K/hour, followed by a more gradual cooling lasting several hours,
with cooling rates of the order of $50$ K/hour. These appear to be the right
conditions for chondrule formation, which is one of the reasons why this
model is one of the favored models of chondrule formation nowadays.

In this paper we revisit this shock-induced chondrule formation model. Our
aim is to investigate the role of the downstream boundary condition and the
role of sideways radiative cooling.

If the shock is a large scale shock, e.g.\ due to a global gravitational
instability, then on a small scale (the scale of several radiative mean free
paths $\lambda_{\mathrm{free}}$) the shock can be modeled as a infinite 1-D
radiation hydrodynamic flow problem. 
The ``infinite 1-D'' in this context means that the 3-D geometry of the
problem only becomes important on scales $\gg \lambda_{\mathrm{free}}$, so
that on scales $\sim \lambda_{\mathrm{free}}$ a 1-D geometry can be safely
assumed where the pre-shock boundary is set at $x=-\infty$ and the post-shock
boundary is set at $x=+\infty$.
With $\lambda_{\mathrm{free}} = \left(
\kappa\rho \right)^{-1}$ the pressure scale height of a typical minimum mass
solar nebula is several hundred times larger than the optical mean free path.
The shock is assumed to be stationary in the
$x$-coordinate system, i.e.\ the coordinates move along with the shock and the
shock is always at $x=0$. In such an infinite 1-D system the full shock
structure can be reconstructed when the physical variables at the upsteam
boundary $x=-\infty$ are all given. No downsteam boundaries at $x=+\infty$
need to be given. In fact, the physical variables at $x=+\infty$ follow
uniquely by demanding that the mass-, momentum- and energy-flux at $x=+\infty$
equals those of $x=-\infty$, but with subsonic gas velocity at $x=+\infty$.
This gives a {\em global} Rankine-Hugoniot condition including all the
radiative and dust physics. Note that right at the shock at $x=0$ the jump in
the gas variables is given by a {\em local} Rankine-Hugoniot condition in
which only the gas fluxes on both sides are set equal. We will work out these
shock models in section \ref{sectionModel} and show that after the temperature
spike they lack a slower few-hour cooling phase. Instead, they stay at a
constant temperature, i.e.\ the temperature in accordance with the global
Rankine-Hugoniot condition. We will investigate in section \ref{sectionStandardShock} whether the chondrule peak
temperature can be made high enough for melting while keeping the post-spike
temperature low enough to retain volatile elements. According to \citet{fedkin12}
and \citet{yu96} the high-temperature phase should be of the order tens of minutes rather
than of hours. From textural constraints, \citet{hewins05} and \citet{desch12}
conclude that the cooling rates have to be of the order of 10$^1$ -- 10$^3$ K/hour.

Since the infinite 1-D shock solution is a geometric approximation we will
implement effects of sideways cooling (i.e., from top and bottom of the disk)
in section \ref{sectionVertLoss} to improve the
realism of the model. This will re-introduce the slow cooling phase after the
temperature spike, but we find that this slow cooling phase is of the order
of weeks/months/years rather than hours, because 2-D/3-D radiative diffusion
will take place on $x$-scales of the same order as the vertical scale height
of the disk.

Finally we discuss in section \ref{sectionOptThin} two versions of the shock-induced chondrule formation
where the cooling time can be rapid. One is a locally
induced one, for instance due to a supersonic planetesimal bow shock
\citep{hood98,morris12}. The other is if the disk is fully optically thin, allowing the post-shock
material to cool straight to infinity. 

\section{The Model}
\label{sectionModel}

Our one-dimensional shock model is built on the work by \citet{desch02}. We
used their approach, generalized it to arbitrarely many gas species, particle
populations and chemical reactions and modified it where we think
modifications or corrections are needed.

In this model we assume that all the parameters (densities, temperatures,
velocities, etc.) only change along one direction, the $x$-direction.
Therefore, the model consists of infinitely extended, plane-parallel layers
of constant temperatures, velocities and densities.

\subsection{Radiative Transfer}
\label{subSectionRadTrans}

Even though the model is one-dimensional, we must allow the photons to move
into all three directions. But fortunately, because of the plane-parallel
assumption the radiative transfer equations here can be vastly simplified.

\begin{figure}
  \includegraphics[width=\linewidth]{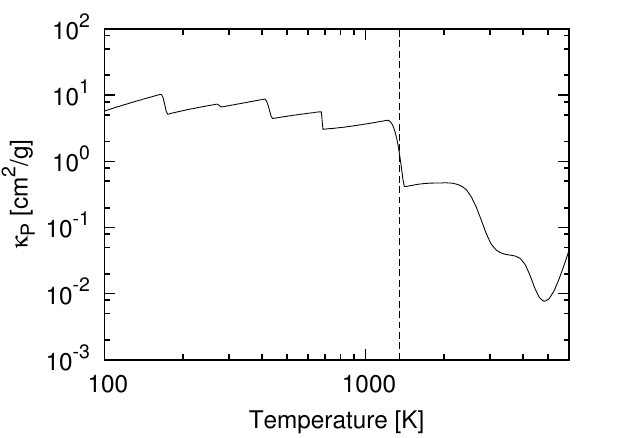}
  \caption{Planck-mean opacity from \citet{semenov03}. At $\sim 1350$ K (vertical line) the
      fine-grained dust associated with the gas gets evaporated. Therefore,
      the opacity drops by one order of magnitude at this temperature. This
      is responsible for the ``opacity knee'' in the pre-heating phase, where
      the gas becomes optically thin.}
  \label{gfxPlanckMeanOpacity}
\end{figure}

In this case the optical depth $\tau$ is a monotonic increasing function of
the distance from the post-shock boundary and  therefore a meassure of the 
position $x$ inside the computational domain:
\begin{equation}
  \tau \left( x \right) = \int\limits_x^{x_\mathrm{post}} \alpha \mathrm{d}x,
\end{equation}
where $x_\mathrm{post}$ is the location of the post-shock boundary. This
means the optical depth increases from $\tau \left( x_\mathrm{post} \right)
= 0$ at the post-shock boundary to a maximum value of $\tau \left(
x_\mathrm{pre} \right) = \int\limits_{x_\mathrm{pre}}^{x_\mathrm{post}}
\alpha \mathrm{d}x \equiv \tau_\mathrm{max}$ at the pre-shock boundary.
$\alpha$ is the absorption coefficient, which is the sum of the
contribution of the gas and the particles:
\begin{equation}
  \begin{split}
    \alpha &= \alpha_{\mathrm{g}} + \alpha_{\mathrm{p}} \\
    &= \rho_{\mathrm{g}} \kappa_\mathrm{P} +  n\pi a^2 \varepsilon .
  \end{split}
\end{equation}
The absoption coefficient of the gas is the product of the gas mass density 
$\rho_{\mathrm{g}}$ and the temperature-dependent Planck-mean opacity
$\kappa_\mathrm{P}$, which we took from \citet{semenov03} (see figure
\ref{gfxPlanckMeanOpacity}).
The absorption coefficient of the particles is the product of their number
densities $n$, their geometrical cross-section and their absoprtion efficiency
$\varepsilon$, which we adopted from \citet{desch02} as
\begin{equation}
  \varepsilon = 0.8 \times \mathrm{min} \left[ 1,\ \left( \frac{a}{2\ 
      \mu\mathrm{m}}
  \right) \right],
\end{equation}
where $a$ is the particle radius.

To calculate the thermal histories of the particles one has to calculate the
mean intensity $J_\mathrm{rad} \left( \tau \right)$ at every position $\tau$
Our radiative transfer
is grey, i.e. wavelength independent. The mean intensity is defined as the
average intensity $I$ per solid angle coming from all directions
\begin{equation}
  J_\mathrm{rad} \left( \tau \right) = \frac{1}{4\pi} \int\limits_\Omega I
      \left( \tau, \mu \right) \mathrm{d}\Omega.
\end{equation}
In the plane-parallel assumption the intensity depends only on the position
$\tau$ and the angle $\theta$ between the incoming ray and the $x$-axis. $\mu$
is defined as $\mu = \cos\theta$.
Therefore, the mean intensity can be simplified to
\begin{equation}
  \begin{split}
    J_\mathrm{rad} \left( \tau \right) = &\frac{I_\mathrm{pre}}{2} E_2 \left(
	\tau_\mathrm{max} - \tau \right) + \frac{I_\mathrm{post}}{2} E_2
	\left( \tau \right) \\
    &+ \frac{1}{2} \int\limits_0^{\tau_\mathrm{max}} S \left( \tau' \right)
	E_1 \left( \left| \tau' - \tau \right| \right) \mathrm{d}\tau',
  \end{split}
\end{equation}
by using the exponential integrals
\begin{equation}
  E_n \left( x \right) = \int\limits_1^\infty \frac{e^{-xt}}{t^n}\mathrm{d}t,
\end{equation}
\citep[see][]{mihalas99}.
$I_\mathrm{pre} = \frac{\sigma}{\pi}T_\mathrm{pre}^4$ and $I_\mathrm{post} =
\frac{\sigma}{\pi}T_\mathrm{post}^4$ are the incoming radiations from both
boundaries. The source function $S\left( \tau \right)$ is defined as
\begin{equation}
  S = \frac{\rho_\mathrm{g} \kappa_\mathrm{P} B\left( T_\mathrm{g} \right) +
      n \pi a^2 \varepsilon B\left( T \right)}{\rho_\mathrm{g}
      \kappa_\mathrm{P} + n \pi a^2 \varepsilon},
\end{equation}
which are the wave-length integrated Plank functions $B \left( T \right) =
\left( \sigma / \pi \right) T^4$ of the gas and the
particles at their given temperatures weighted by their respective emission
efficiencies, which are the same as their absoprtion efficiencies according
Kirchoff's law. $\sigma$ is the Stefan-Boltzmann constant.

The radiative flux $F_\mathrm{rad}$ is defined as the net flux of radiative
energy in $x$-direction
\begin{equation}
  \begin{split}
    F_\mathrm{rad} &= -\int\limits_\Omega I \left( \tau,\mu \right) \cos\theta
	\mathrm{d}\Omega \\
    &= -2\pi\int\limits_{-1}^1 I \left( \tau,\mu \right) \mu	\mathrm{d}\mu.
  \end{split}  
\end{equation}
Later, we need the derivative of the radiative flux with respect to the
$x$-direction. In the plane-parallel assumption this is simply
\begin{equation}
  \begin{split}
    \frac{\partial F_\mathrm{rad}}{\partial x} = &-4\pi\alpha \left(
	J_\mathrm{rad} - S \right) \\
    = &-4\pi\rho_\mathrm{g}\kappa_\mathrm{P} \left( J_\mathrm{rad} -
	\frac{\sigma T_\mathrm{g}^4}{\pi} \right) \\
    &-4\pi\sum\limits_{j=1}^{N_J} n_j \pi a_j^2 \varepsilon_j \left(
	J_\mathrm{rad} - \frac{\sigma T_\mathrm{j}^4}{\pi} \right),
  \end{split}
  \label{eqnFrad}
\end{equation}
where we now included $N_J$ different particle populations. Therefore, we have
to take the sum over all populations.

\subsection{Hydrodynamics}
\label{subSectionHydDyn}

To calculate the evolution of the gas and particle parameters we used the
one-dimensional stationary Euler equations, which are
\begin{align}
  \partial_x \left[ \rho V \right] &= S_\rho 
    \label{eqnEulerEquationMass} \\
  \partial_x \left[ \rho V^2 + P \right] &= S_p 
    \label{eqnEulerEquationMomentum} \\
  \partial_x \left[ \left( \rho e_\mathrm{tot} + P \right) V \right] &= S_e
    \label{eqnEulerEquationEnergy},
\end{align}
with the total mass density $\rho$, the velocity $V$, the pressure $P$ and
kinetic and thermal energy density $e_\mathrm{tot}$. These equations are simply
conservation laws of mass, momentum and energy respectively. S$_\rho$,
S$_p$ and S$_e$ are the source terms of the designated quantities.

\subsubsection{Particle dynamics}
\label{subsubsectionPartDyn}

In this model we have $N_J$ different particle populations with radius $a_j$,
temperature $T_j$, velocity $V_j$ and number density $n_j$. The particles
are not allowed to evaporate completely. This means the continuity equation
(\ref{eqnEulerEquationMass}) states
\begin{equation}
  \partial_x \left[ n_j V_j \right] = 0.
  \label{eqnParticlesContinuity}
\end{equation}

The particles are accelerated or decelerated by the drag force
$F_{\mathrm{drag},j}$ acting on population $j$. The drag force is given by
\begin{equation}
  F_{\textrm{drag},j} = -\pi a_j^2 \rho_\mathrm{g} \frac{C_{\mathrm{D},j}}{2}
      \left| V_j - V_\mathrm{g} \right| \left( V_j - V_\mathrm{g} \right)
  \label{eqnDragForce}
\end{equation}
\citep{gombosi86}, with the gas' mass density $\rho_\mathrm{g}$ and the drag
coefficient $C_{\mathrm{D},j}$ of population $j$ (see
\ref{sectionAppendixDragForce} for futher details). If the gas velocity is
higher than the particle velocity, the particles get accelerated and vice
versa. With that, the force equation of the particles is
\begin{equation}
  m_j \frac{\mathrm{D}}{\mathrm{D} t} V_j = F_{\mathrm{drag},j},
  \label{eqnParticlesForce}
\end{equation}
with the particles' mass $m_j$ and the comoving derivative
$\mathrm{D} / \mathrm{D}t$, which is defined as
\begin{equation}
  \frac{\mathrm{D}}{\mathrm{D}t} \equiv \frac{\partial}{\partial t} +
      \vec V_j \cdot \vec\nabla = V_j \frac{\partial}{\partial x}.
  \label{eqnComovingDerivative}
\end{equation}
The last equality is for the one-dimensional, stationary case. Using this,
the force equation (\ref{eqnParticlesForce}) yields
\begin{equation}
  m_j V_j \frac{\partial}{\partial x} V_j = F_{\mathrm{drag},j}.
  \label{eqnParticlesForce2}
\end{equation}

The energy budget of the particles is given by the balance between frictional
heating by the gas drag, thermal contact with the gas and radiative heating
on one hand and radiative cooling on the other hand. The effects of
frictional heating and thermal contact with the gas are combined into a
single heating rate $q_j$ per unit surface area of the particle given by
\begin{equation}
  q_j = \rho_\mathrm{g} C_{\mathrm{H},j} \left( T_\mathrm{rec} - T_j \right)
      \left| V_\mathrm{g} - V_j \right|
  \label{eqnFrictionalHeatingRate}
\end{equation}
\citep{gombosi86}, with the heat transfer coefficient $C_{\mathrm{H},j}$ and
the recovery temperature $T_\mathrm{rec}$. In the limit of
$V_\mathrm{g} = V_\mathrm{j}$ there can be still heat exchange if
$T\mathrm{g} \neq T\mathrm{j}$ (see \ref{sectionAppendixDragForce}
for further details).
The radiative heating rate per unit surface area of the paticles is given by
$\varepsilon_j \left( \pi J_\mathrm{rad} - \sigma T_j^4 \right)$, which is
the balance between radiation received by the mean intensity $J_\mathrm{rad}$
and the energy radiated away according the Stefan-Boltzmann law.
Combining the effects of frictional and thermal heating and radiative heating
the net heating rate per unit surface area is $q_j + \varepsilon_j
\left( \pi J_\mathrm{rad} - \sigma T_j^4 \right)$.

\begin{figure}
  \includegraphics[width=\linewidth]{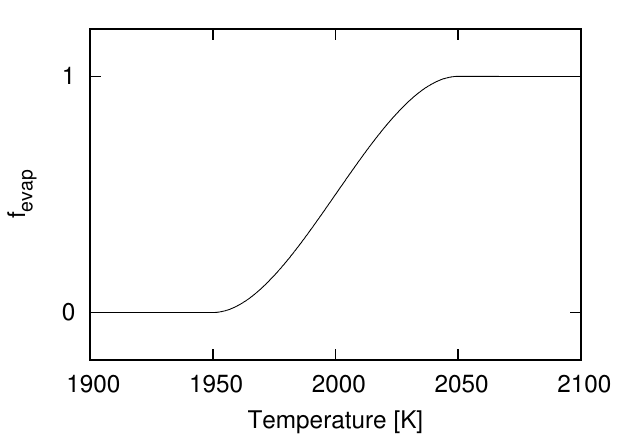}
  \caption{The fraction of evaporation $f_\mathrm{evap}$ yields for a
      smearing of the evaporation temperature of 2000 K within our choice
      of the interval $\Delta T = 100 ~\mathrm{K}$.}
  \label{gfxFevap}
\end{figure}

This net heating rate can raise the particles' temperatures and/or evaporate
them. We assume that the fraction of the net heating rate that goes into
evaporation is $f_{\mathrm{evap},j}$, which is a function of the particle
temperature. $f_{\mathrm{evap},j}$ increases monotonically from 0 to 1 within
a temperature interval $\Delta T$ centered on the evaporation temperature of
2000 K used by \citet{desch02}. We have arbitrarely chosen
$\Delta T = 100 ~\mathrm{K}$ (see figure \ref{gfxFevap}). We had to do this because of
numerical reasons (strong if-statements can make it impossible to estimate
the Jacobian for the implicit integration scheme), but the positive side
effect is that we can account now for the heterogeneity of the particle
material with different evaporation temperatures.

Therefore, the change in the particles' temperature is given by
\begin{equation}
  \begin{split}
    m_j C_{\mathrm{P},j} \frac{\mathrm{D}}{\mathrm{D}t} T_j = &\left(
	1 - f_{\mathrm{evap},j} \right) \times \\
    &\times 4 \pi a_j^2 \left[ q_j + \varepsilon_j \left( \pi J_\mathrm{rad} - \sigma T_j^4
	\right) \right],
  \end{split}
\end{equation}
with the specific heat capacity at constant pressure $C_{\mathrm{P},j}$ of
the particle material. Using again the definition of the comoving derivative
(\ref{eqnComovingDerivative}), the particles' material mass density $\rho_j$
and using $m_j = 4/3 \pi \rho_j a_j^2$ the change in the particles'
temperature can be written as
\begin{equation}
    \frac{\partial}{\partial x} T_j = \frac{3\left( 1 - f_{\mathrm{evap},j}
	\right)}{\rho_j a_j V_j C_{\mathrm{P},j}} \left[ q_j + \varepsilon_j
	\left( \pi J_\mathrm{rad} - \sigma T_j^4 \right) \right].
    \label{eqnParticlesTemperature}
\end{equation}
The other part of the net heating rate, that goes into evaporation, is given
by
\begin{equation}
  \begin{split}
    H_{\mathrm{evap},j}\frac{\mathrm{D}}{\mathrm{D}t} m_j =
	&-f_{\mathrm{evap},j} \times \\
    &\times 4 \pi a_j^2 \left[ q_j + \varepsilon_j \left( \pi J_\mathrm{rad} -
	\sigma T_j^4 \right) \right],
  \end{split}
\end{equation}
with the latent heat of evaporation $H_{\mathrm{evap},j}$ of the particle
material. The negative sign is needed because the heat received here  goes
into shrinking the particles' mass. This yields for the change in the particle radius
\begin{equation}
  \frac{\partial}{\partial t} a_j =
      -\frac{f_{\mathrm{evap},j}}{\rho_j H_{\mathrm{evap},j} V_j} \left[ q_j
      + \varepsilon_j \left( \pi J_\mathrm{rad} - \sigma T_j^4 \right) \right].
  \label{eqnParticlesRadius}
\end{equation}

If the net heating rate is negative then $f_{\mathrm{evap},j}$ is set to
zero to avoid artificial condensation in equation (\ref{eqnParticlesRadius}).
We neglect condensation and nucleation here because we assume it takes place on much longer
timescales than considered here. We want to point out that this implementation
of evaporation is not the correct physical treatment. To do it in an correct
way one has to integrate the Hertz-Knudsen equation and possibly a nucleation model, which makes the
whole system even more complex. We also want to point out that in our
conclusive simulations, the particle temperature is always safely below the
evaporation temperature.

With the continuity equation of the particles (\ref{eqnParticlesContinuity})
and the change in the particle velocity (\ref{eqnParticlesForce2}), the
change in the particles' number densities can be calculated
\begin{equation}
  \frac{\partial}{\partial x} n_j = -\frac{n_j}{V_j}
      \frac{\partial}{\partial x} V_j.
  \label{eqnParticlesNumberDensity}
\end{equation}

With equations (\ref{eqnParticlesForce2}), (\ref{eqnParticlesTemperature}),
(\ref{eqnParticlesRadius}) and (\ref{eqnParticlesNumberDensity}) all the
particle parameters can be calculated throughout the computational domain
knowing the gas parameters.

\subsubsection{Gas Dynamics}
\label{subsubsectionGasDyn}

The gas consists of $N_I$ different gas species, which are assumed to be
well-coupled and share the same temperature $T_g$ and velocity $V_g$. Each
species $i$ has a number density $n_i$ (do not confuse this with $n_j$ of
the particles) and a mean molecular weight $m_i$.

The model also includes $N_K$ different chemical reactions with their
respective net reaction rates $R_k$ of reaction $k$, which are the number of
reactions per unit time per unit volume. In chemical equilibrium the net
reactions rates are equal to zero (which does not mean that no chemical
reaction takes place).

Every reaction costs or sets free energy. The definition is
such, that a positive reaction rate $R_k$ sets free the energy $e_k$. That
means that the total energy per unit time and unit volume set free by
reaction $k$ is $R_ke_k$.
With chemistry the continuity equations of the different gas species are
\begin{equation}
  \frac{\partial}{\partial x} \left( n_iV_\mathrm{g} \right) =
      \sum_{k=0}^{N_K} R_{i,k},
  \label{eqnGasContinuity}
\end{equation}
where $R_{i,k}$ is the creation rate of gas species $i$ due to reaction $k$
(i.e., the change in number density $n_i$ due to reaction k).
The $k=0$ component of the creation rates $R_{i,0}$ accounts for changes due
to non-chemical processes, e.g. evaporation since the mass of the evaporated
material has to be added to the gas.

The total mass loss per unit time and unit volume of all particle populations
is
\begin{equation}
  \sum_{j=1}^{N_J} n_j \frac{\mathrm{D}}{\mathrm{D} t} m_j = \sum_{j=1}^{N_J} n_j
      V_j 4 \pi \rho_j a_j^2 \frac{\partial}{\partial x} a_j.
\end{equation}
This mass has to be added to the different gas species via the $k=0$
component of their creation rates
\begin{equation}
  R_{i,0} = -\frac{\xi_{\mathrm{evap},i}}{m_i} \sum_{j=1}^{N_J} n_j V_j 4
      \pi \rho_j a_j^2 \frac{\partial}{\partial x} a_j,
\end{equation}
where $\xi_{\mathrm{evap},i}$ is the fraction of the evaporated mass added
to gas species $i$. For mass conservation $\sum\limits_i
\xi_{\mathrm{evap},i} = 1$ has to hold.

To close the system of equations one has to solve for the gas temperature
$T_\mathrm{g}$, the gas velocity $V_\mathrm{g}$ and the number densities
$n_i$. To do so we use Euler's momentum equation
(\ref{eqnEulerEquationMomentum}) and sum up the total momentum of the gas and
the particles
\begin{equation}
  \begin{split}
    \frac{\partial}{\partial x} \left[ \sum_{i=1}^{N_I} \left(
	\rho_i V_\mathrm{g}^2 + n_i k_\mathrm{B} T_\mathrm{g} \right) +
	\sum_{j=1}^{N_J} n_j m_j V_j^2 \right] = 0.
  \end{split}
  \label{eqnTotalMomentum}
\end{equation}
The first term in the sum over $i$ is the momentum carried by the gas, the
second term is the pressure according the ideal gas law. The sum over $j$
refers to the momentum carried by the particles. Since the particles are
pressureless there is no pressure term. This formula can be manipulated to
get to
\begin{equation}
  \begin{split}
    & \left( \rho_\mathrm{g} V_\mathrm{g} - \frac{k_\mathrm{B}
	T_\mathrm{g}}{V_\mathrm{g}} \sum_{i=1}^{N_I} n_i \right)
	\frac{\partial}{\partial x} V_\mathrm{g} + \left( k_\mathrm{B}
	\sum_{i=1}^{N_I} n_i \right) \frac{\partial}{\partial x} T_\mathrm{g} \\
    = & - \sum_{j=1}^{N_J} n_j \left( F_{\mathrm{drag},j} + 4 \pi a_j^2
	\rho_j V_j^2 \frac{\partial}{\partial x} a_j \right) \\
    & - \sum_{i=1}^{N_I} \sum_{k=0}^{N_K} R_{i,k} \left( V_\mathrm{g} m_i +
	\frac{k_\mathrm{B} T_\mathrm{g}}{V_\mathrm{g}} \right)
  \end{split}
  \label{eqnCoupled1}
\end{equation}
(see \ref{sectionAppendixHydDynCalc} for detailed calculations).

The same kind of calculations can be done for Euler's energy equation
(\ref{eqnEulerEquationEnergy})
\begin{equation}
  \begin{split}
    &\frac{\partial}{\partial x} \left[ \sum_{i=1}^{N_I} n_i V_\mathrm{g} \left(
	\frac{m_i V_\mathrm{g}^2}{2} + \frac{f_i + 2}{2} k_\mathrm{B}
	T_\mathrm{g} \right) \right. \\
    & \left. + \sum_{j=1}^{N_J} n_j m_j V_j \left( \frac{V_j^2}{2} +
	C_{\mathrm{P},j} T_j \right) + F_\mathrm{rad} \right] \\
    = & \sum_{k=0}^{N_K} R_k e_k.
  \end{split}
  \label{eqnTotalEnergy}
\end{equation}
The first term in the sum over $i$ is the kinetic energy carried by the gas,
whereas the second term is the sum of the internal energy of the gas and the
pressure term according the ideal gas law. $f_i$ is the number of degrees of
freedom of gas species $i$. The first term in the sum over $j$ is the kinetic
energy carried by the particles, the second term is the thermal energy of the
particles. In addition to that the radiative flux $F_\mathrm{rad}$ has to be
included. The right hand side is not equal to zero, because energy can be
created or consumed by chemical reactions. This equation can be further
manipulated to
\begin{equation}
  \begin{split}
    & V_\mathrm{g}^2 \sum_{i=1}^{N_I} n_i m_i \frac{\partial}{\partial x}
	V_\mathrm{g} + k_\mathrm{B} V_\mathrm{g} \sum_{i=1}^{N_I} n_i
	\frac{f_i + 2}{2} \frac{\partial}{\partial x} T_\mathrm{g} \\
    = \quad & 4 \pi \rho_\mathrm{g} \kappa \left( J_\mathrm{rad} - \frac{\sigma
	T_\mathrm{g}^4}{\pi} \right) + 4 \pi \sum_{j=1}^{N_J} n_j \pi a_j^2
	\varepsilon_j \left( J_\mathrm{rad} - \frac{\sigma T_j^4}{\pi} \right) \\
    & - \sum_{i=1}^{N_I} \sum_{k=0}^{N_K} R_{i,k} \left( \frac{m_i V_\mathrm{g}^2}{2}
	+ k_\mathrm{B} T_\mathrm{g} \frac{f_i + 2}{f_i} \right) \\
    & - \sum_{j=1}^{N_J} n_j V_j \left( F_{\mathrm{drag},j} + 2 \pi V_j^2
	\rho_j a_j^2 \frac{\partial}{\partial x} a_j \right) \\
    & - \sum_{j=1}^{N_J} n_j V_j m_j C_{\mathrm{P},j} \frac{\partial}{\partial x}
	T_j \\
    & - 4 \pi \sum_{j=1}^{N_J} n_j V_j C_{\mathrm{P},j} T_j a_j^2 \rho_j
	\frac{\partial}{\partial x} a_j \\
    & + \sum_{k=1}^{N_K} R_k e_k
  \end{split}
  \label{eqnCoupled2}
\end{equation}
(see \ref{sectionAppendixHydDynCalc} for detailed calculations).

Equations (\ref{eqnCoupled1}) and (\ref{eqnCoupled2}) are a set of two
coupled differential equations, which can be solved for
$\frac{\partial}{\partial x} V_\mathrm{g}$ and $\frac{\partial}{\partial x}
T_\mathrm{g}$. Together with the continuity equation (\ref{eqnGasContinuity})
of the gas and the particle differential equations (\ref{eqnParticlesForce2}),
(\ref{eqnParticlesTemperature}), (\ref{eqnParticlesRadius}) and
(\ref{eqnParticlesNumberDensity}) this closes the system of equations.

Setting $N_I=4$ and using the gas species $\mathrm{H}$, $\mathrm{H_2}$,
$\mathrm{He}$ $\mathrm{and}$ $\mathrm{SiO}$ this reduces in principle to the
model of \citet{desch02}, but correcting for some sign errors and
implementing the smooth transition of the evaporation temperature.

\subsection{Numerical Method}
\label{subSectionNumMeth}

The system of equations (\ref{eqnParticlesForce2}),
(\ref{eqnParticlesTemperature}), (\ref{eqnParticlesRadius}),
(\ref{eqnParticlesNumberDensity}), (\ref{eqnGasContinuity}),
(\ref{eqnCoupled1}) and (\ref{eqnCoupled2}) is extremely stiff. Therefore it
would require a very small step size and a very large number of grid points to
numerically integrate it with an explicit integration scheme. We used here the implicit integrator \emph{DVODE} \citep{brown89}
to integrate the equations simultaneously through the computational domain.

We performed our calculations in the comoving frame of the shock front,
which is set to be at $x=0$. At the shock front the gas parameters are
changed according the Rankine-Hugoniot jump conditions, while the particle
parameters remain unchanged.

After every complete integration the radiative transfer calculations have to
be done again with the new particle and gas parameters. Then the integration
is repeated with these new radiative parameters. This is repeatedly done
until convergence is reached.

Another crucial point is the calculation of the post-shock boundary
temperature $T_\mathrm{post}$, which is needed for the radiative transfer
calculation. \citet{desch02} calculated $T_\mathrm{post}$ by using incorrect jump
conditions adopted from \citet{hood91}. This was corrected by \citet{morris10}
by calculating their own jump conditions. They found post-shock temperatures
on the order of $>1300 ~\mathrm{K}$ and concluded that chondrule formation
is not possible in strictly one-dimensional models. Since disks are not
one-dimensional objects they will eventually cool by radiation. Therefore
\citet{morris10} loosened the one-dimensional assumption by setting the
post-shock temperature to the pre-shock temperature
$T_\mathrm{post} = T_\mathrm{pre}$. 
This approach raises a problem, however, since it forces the gas and dust to radiatively cool through the
downstream boundary. Since this downstream boundary is not a physical boundary, but just a
computational boundary, this does not appear to be justified.
The cooling is then dependent on the distance between shock front and
post-shock boundary.

In our model we do not set the post-shock temperature a-priori since in a
one-dimensional stationary model all downstream parameters are completely set
by the upstream conditions. Therefore we perform right before
the first iteration an additional integration without radiation. The
temperature reached there at the post-shock boundary is then used as a first
approximation of the post-shock temperature. After every further iteration,
we check the radiative flux through the post-shock boundary. If the flux is
positive i.e., the final temperature is higher than the post-shock
temperature, we increase $T_\mathrm{post}$ slightly and vice versa. If
convergence is reached then the radiative flux at the post-shock boundary is
equal to zero. We want to point out that $F_\mathrm{rad} = 0$ is only true
at the boundaries, which have to be at large enough $|x|$.

The post-shock temperatures we found by this approach are even higher than
those calculated by \citet{morris10}. We think this is due to some
approximations performed in deriving their jump conditions. Later in section
\ref{sectionResults} we introduce a method to perform simulations with a
vertical energy loss.

\section{Results}
\label{sectionResults}

For reasons of comparison we used the same input parameters as 
\citet{desch02}, which we want to repeat here.

We have one particle population with initial radius of \mbox{300 $\mu$m} and
an material mass density of \mbox{3.3 g/cm$^3$}. Their initial number density
can then be calculated by assuming a gas density of \mbox{$10^{-9}$
g/cm$^3$}, a dust-to-gas ratio of 0.005 and assuming that \mbox{75 \%} of the dust
mass are in the chondrule precursors. The particles have a heat capacity of
\mbox{$C_\mathrm{P} = 10^{7}$ erg/g/K}. Between 1400 K and 1820 K melting
takes place. This results in an effective heat capacity of
\mbox{$C_\mathrm{P} = 2.19 \cdot 10^{7}$ erg/g/K} within this temperature
interval \citep[see][and references therein for detailed descriptions]
{desch02}. The latent heat of evaporations is
$H_\mathrm{evap} = 1.1 \cdot 10^{11}$ erg/g.

The intial temperature of both gas and dust is \mbox{300 K}. The gas consists
of four species: atomic hydrogen, molecular hydrogen, helium and silicon
monoxide (SiO). The only chemical reaction we consider is hydrogen
dissociation and recombination, which consumes 4.48 eV for every H$_2$
molecule that breaks apart. We used the reaction rates given in \citet{desch02} adopted
from \citet{cherchneff92}. The reaction can also go in reverse direction; then setting
free energy. If the gas exceeds 1350 K, then the dust
associated with it evaporates. The mass of the evaporated dust gets added to
the silicon monoxide.

\subsection{The standard shock}
\label{sectionStandardShock}

\begin{figure*}
  \begin{tabular}{cc}
    \includegraphics[width=0.5\linewidth]{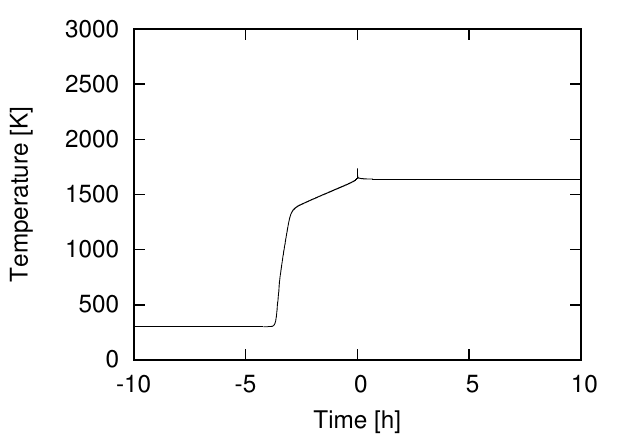} &
    \includegraphics[width=0.5\linewidth]{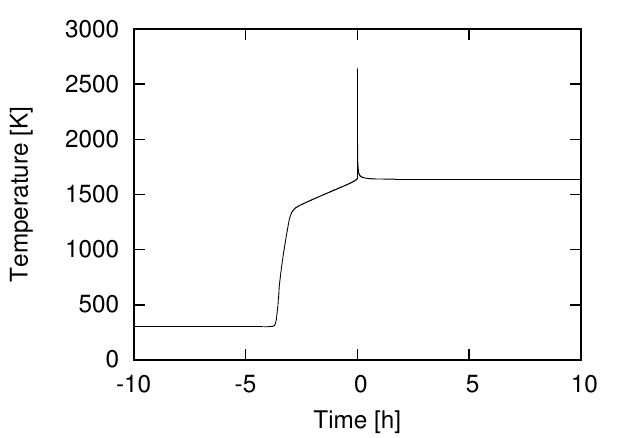}
  \end{tabular}
  \caption{Left: Particle temperature in a shock with shock speed
      $V_\mathrm{shock}=6.5$ km/s. Right: Same as left but with gas
      temperature. See text for further description.}
  \label{gfxStandardShock}
\end{figure*}

A standard shock with a shock speed of $V_\mathrm{shock} = 6.5$ km/s is
shown in figure \ref{gfxStandardShock}. Most of the time the gas and the
particles are well-coupled and share the same temperature. Already 3-4 hours
before the shock front, the particles receive radiation from the hot gas and
particles behind the shock. The temperature increases rapidly until roughly
1350 K. From that point on the temperature only slightly increases towards
the shock front. This change in slope is related to the opacity (see figure 
\ref{gfxPlanckMeanOpacity}). As the gas reaches temperatures of $\sim$ 1350 K
the fine grained dust associated with the gas gets evaporated and the opacity
drops by one order of magnitude. From that point on the opacity is purely
caused by molecular lines. Therefore the pre-heating region can be
sub-divided into an optically thick and an optically thin region.
It is important that the opacity is not set to
zero because then the gas could not actively cool anymore. The only chance
to lose energy would then be thermal contact with the particles. This would
artificially slow down the cooling process.

At the position of the shock front the parameters of the gas only are changed
(i.e., without radiation or dust) according the Rankine-Hugoniot jump
conditions wheareas the particle parameters remain unchanged. Therefore, the
gas reaches temperatures as high as 2700 K. The particles suddenly find
themselves surrounded by gas that is much hotter than before the shock and that
has velocities much smaller than those of the particles. The particles are now
heated up by thermal contact with the hot gas and by frictional heating.
But since the radiative cooling into the pre-shock region
is relatively effective, both the gas and the particles quickly reach an equilibrium
post-shock temperature before the particles are able to adapt to the high gas
temperatures. In that way the maximum temperature the particles reach is
$\sim 1000$ K smaller than the maximal gas temperature.

The constant post-shock equilibrium temperature is a consequence of the
one-dimensionality of the model. As soon as the gas and the particles are a
few optical depths behind the shock front there is no way for them to lose
energy into the pre-shock region by radiation. Since the model consists of
infinitely extended plane-parallel layers, they can not cool in lateral
directions.

Therefore our model has in principle two different jump conditions: one is
local at the position of the shock front, where only the gas parameters are
changed according the Rankine-Hugoniot jump conditions. The other \emph{global}
jump condition forces the gas and particles to be at rather high post-shock
temperatures after the relaxation of the temperature spike at the shock front.

From now on the maximum temperature the particles experience in the spike is
referred to as peak temperature $T_\mathrm{peak}$, whereas the equilibrium
post-shock temperature is $T_\mathrm{post}$.

The results for different shock speeds look similar. The lower
the shock speed, the lower $T_\mathrm{peak}$ and $T_\mathrm{post}$ and the
smaller the pre-heating region. Only if the shock speed is too low for the gas
to reach 1350 K, then there's also a lack of the ``opacity knee'' noticable.

The chemical reaction of hydrogen dissociation works as energy sink in this
case. Instead of raising the temperature, energy is consumed in breaking
molecular bonds. Simulations without chemical reactions show an overall
increase in temperature downstream by $\sim 100$ K. We want to point out
that only a small fraction of the H$_2$ is dissociated.
This demonstrates that it is very
important to include chemistry in such simulations: the temperature deviation
can make the difference between losing particles via evaporation or not.
Whether or not other chemical reactions are equaly important has to be
investigated.

\begin{figure}
  \includegraphics[width=\linewidth]{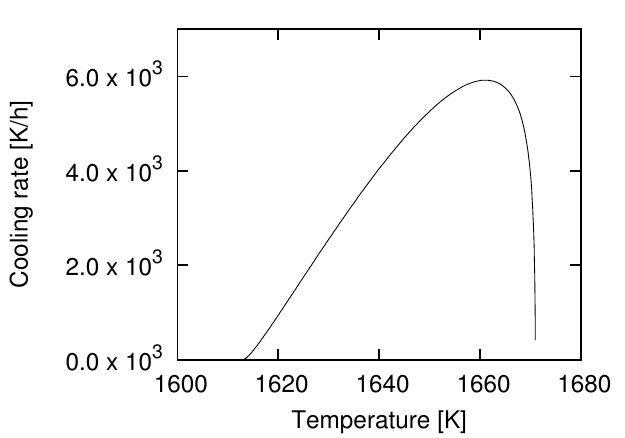}
  \caption{Cooling rates of the particles at particle temperatures T $> 1500$ K in the standard
      shock}
  \label{gfxStandardCoolingRates}
\end{figure}

Figure \ref{gfxStandardCoolingRates} shows the cooling rates of the particles
in the standard shock at temperatures T $> 1500$ K which is the important
temperature regime for crystallization. As seen here the peak of the cooling
rates is at the upper limit of what is allowed by experimental constraints. The
standard shock itself is too weak to completely melt the particles. In addition
to that the cooling rates drop to zero at already ~1600 K. This is discussed
later in this section.

\begin{figure}
  \includegraphics[width=\linewidth]{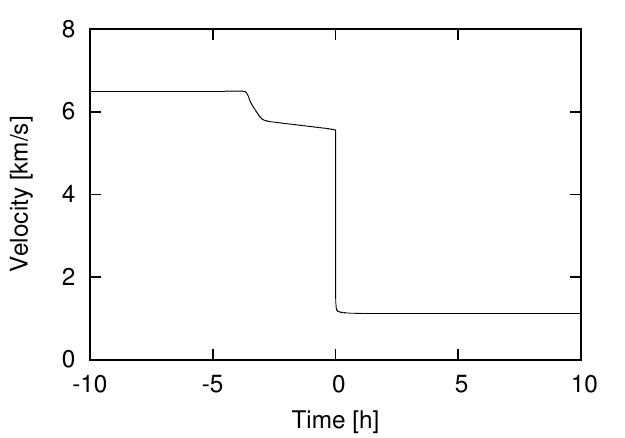}
  \caption{Velocity of the particles.}
  \label{gfxStandardVelocity}
\end{figure}

Another interesting feature is shown in figure \ref{gfxStandardVelocity}: the
velocity of the particles in the simulation. As soon as the pre-heating sets
in the particles' velocity decreases because the gas velocity decreases. From
initially $6.5$ km/s to roughly $5.5$ km/s shortly before the shock front.
After the gas velocity is changed according the Rankine-Hugoniot conditions
at the shock front the particles are rapidly decellerated by the drag force
within minutes to $\sim 1.0$ km/s.

The Mach number of the gas is initially of the order of 5. But due to the
increase in temperature and decrease in velocity it is only roughly 2 at the
position of the shock, where the Mach number is applied in the
Rankine-Hugoniot conditions.

\begin{figure}
  \includegraphics[width=\linewidth]{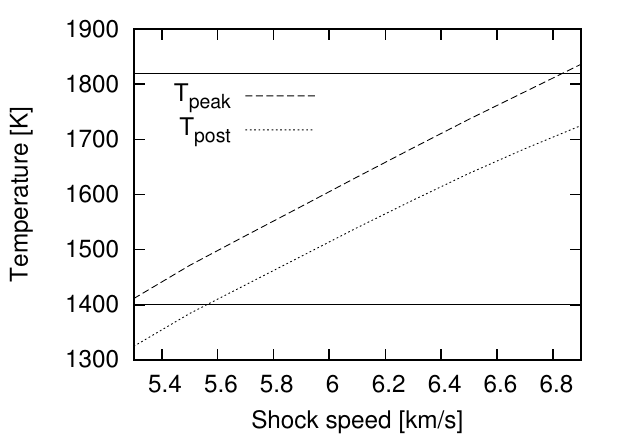}
  \caption{The peak and post-shock temperatures of the standard shock for different shock
      speeds. The upper solid line denotes a temperature of \mbox{1820 K}, which we assume
      the particles have to reach to be completely molten. The lower solid
      line is at \mbox{1400 K}. This is the temperature the particles have
      to reach at least after the shock to retain their volatiles. This is not possible in
      the standard case.}
  \label{gfxTPeakPost}
\end{figure}

\begin{figure*}
  \begin{tabular}{cc}
    \includegraphics[width=0.5\linewidth]{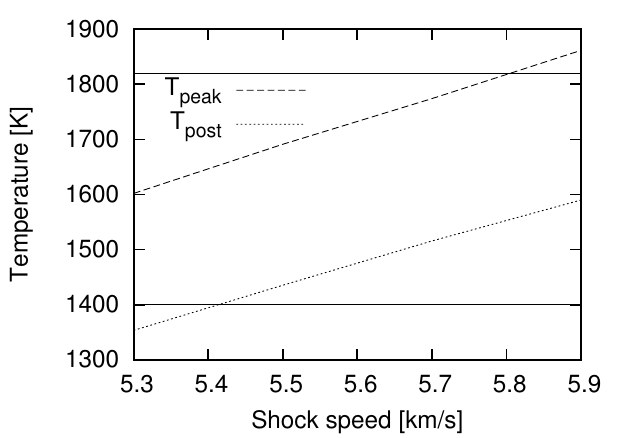} &
    \includegraphics[width=0.5\linewidth]{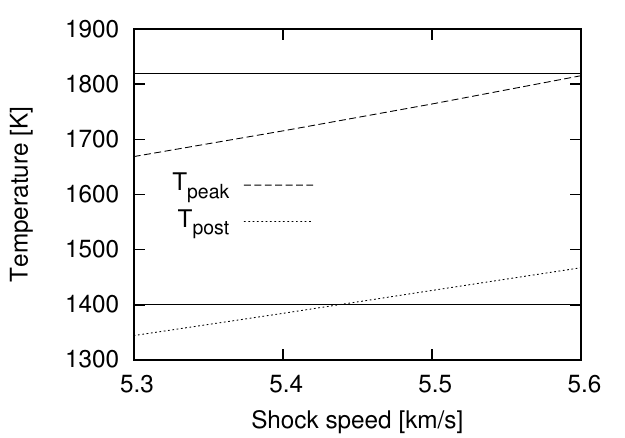}
  \end{tabular}
  \caption{Same as figure \ref{gfxTPeakPost} but with higher gas mass
      densities of 10$^{-8}$ g/cm$^3$ (left) and 10$^{-7}$ g/cm$^3$ (right).}
  \label{gfxTPeakPostHighDens}
\end{figure*}

The problem of chondrule formation within this scenario is summarized in
figure \ref{gfxTPeakPost}. If we assume that the particles have to reach a
temperature of at least 1820 K to be completely molten and at the same time
to cool rapidly back down below at least 1400 K to retain volatile
materials which are observed in them, this is not possible in a
one-dimensional model of an optically thick disk.

If we increase the gas mass density (see figure \ref{gfxTPeakPostHighDens})
the problem improves slightly and the two
lines of peak and post-shock temperature are further apart. But only at
unrealistic densities of $\gtrsim 10^{-6}$ g/cm$^3$ we could have both
requirements fulfilled at the same time. According the disk model by
\citet{bell97} these high midplane densities can only be found at radii of
$\sim 0.001$ AU or closer to the sun, where the temperatures are already on
the order of 10$^3$ K to 10$^4$ K depending on the accretion rate.

But we want to point out that
post-shock temperatures of 1400 K might already be too high to retain the
volatiles. In addition to that the particles are already at high
temperatures in the pre-heating phase for a prolonged time ($\sim$ 3 h for the standard shock,
cf. figure \ref{gfxStandardShock}),
which is already too long to be consistent with chondrule formation \citep{fedkin12}.

\subsection{Vertical energy loss}
\label{sectionVertLoss}
It is clear that a one-dimensional, plane-parallel model does not match the
situation in actual protoplanetary disks, since such disks are not
one-dimensional objects. They have a vertical extent with decreasing
densities at higher altitudes above the midplane. At $R = 3$ AU (the region of
the asteroid belt where most of the chondrules can be found today) the pressure
scale height of a disk around a solar mass star with a midplane temperature of
$T = 300$ K is $H_\mathrm{P} =
\frac{k_\mathrm{B}TR^3}{\mu m_\mathrm{p}GM_{\odot}} \simeq 0.1$ AU, with the
mean molecular weight of the gas $\mu \simeq 2.2$ amu and the proton mass
$m_\mathrm{p}$.

If the disk is heated up by a shock whose propagation direction lies in the
plane of the disk it can cool in vertical direction by radiative diffusion.
But to cool down to the pre-shock temperature the gas and particles have to
travel \emph{at least} a distance comparable to the disk height. This is
because in a vertically optically thick disk the radiative diffusion is a
photon random walk: before it finds its
way up a distance $H_\mathrm{P}$, it has an equal chance of moving a distance
$H_\mathrm{P}$ downstream. Radiative diffusion cooling will thus not create
a steeper temperature gradient in the downstream (in-plane) direction than in
vertical direction where the radiation escapes. A downstream cooling length
of at least $H_\mathrm{P}$ (assuming $H_\mathrm{P} = 0.1$ AU) amounts, with
a gas velocity of 6 km/s, to at least one month.

To estimate this we added an energy loss term to our equations
\begin{equation}
  \frac{\partial}{\partial x} T^4 = - \frac{T^4-T_\mathrm{pre}^4}{L}.
\end{equation}
We transferred this energy loss into a temperature change and included it
into the differential equation of our gas temperature
\begin{equation}
  \frac{\partial}{\partial x} T_\mathrm{g} = \ \cdots \
      - \frac{T_\mathrm{g}^4-T_\mathrm{pre}^4}{4T_\mathrm{g}^3 L},
\end{equation}
where the $\cdots$ denote the terms in the standard equation. The parameter
$L$ is a length scale, which determines the strength of the energy loss. The
larger $L$, the smaller is the energy loss. Therefore $L$ is also a rough
estimate of the vertical extent of the disk assumed. With time
the temperature should approach the pre-shock temperature. \highlight{To get a
definitive answer one has to perform fully three-dimensional radiative transfer
calculations. But this is beyond the scope of this paper.}

\begin{figure}
  \includegraphics[width=\linewidth]{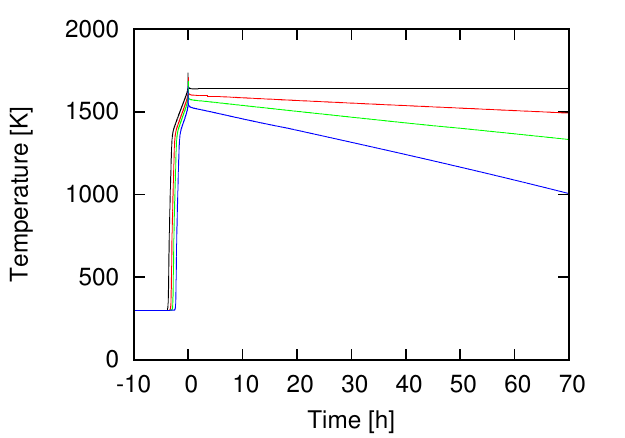}
  \caption{The standard shock (black) compared to simulations with a vertical energy
      loss with different loss parameters $L$ of \mbox{4 \%} (red),
      \mbox{2 \%} (green) and \mbox{1 \%} (blue) of the pressure scale height
      $H_\mathrm{P} = 0.1$ AU at 3 AU. The shock speed is 6.5 km/s in all
      cases.}
  \label{gfxLengthScales}
\end{figure}

We have done this for different length scale parameters essentially
simulating disks in which all the dust has settled into a very thin midplane
layer.
The results are compared to the standard case, which corresponds to
$L = \infty$, in figure \ref{gfxLengthScales}. As seen here, the higher the
energy loss (the lower the length scale $L$), the smaller are the pre-shock
regions and the peak temperatures. The length scale parameters chosen here
are extremely small, only a few percent of the disk's pressure scale height.
It is very questionable if such disks exist. And even in those thin disks the
particles are cooked at temperature above 1400 K for hours.
The shock speed here is not even enough to completely melt the
particles. With the vertical energy loss the particles still spend
2-3 hours in the high-temperature pre-shock phase.
At higher shock speeds and therefore temperatures the situation is
even worse.

\subsection{Optically thin case}
\label{sectionOptThin}
To investigate an optically thin case, where the particles can freely lose
energy by radiation, we did another run where we set the mean intensity to
$J_\mathrm{rad}\left( \tau \right) = \frac{\sigma}{\pi}T_\mathrm{pre}^4$ at every position.
This means the gas and the particles are always in a radiation field with an
ambient temperature of $T_\mathrm{pre}$. At the shock front the gas parameters
are changed as usual and the particles can adapt to it. This could correspond
for example to bow shocks created by planetesimals on eccentric orbits
\citep{hood98}. Here we assume the shocked volume of the disk is small
compared to the unshocked medium such that the particles mostly see the radiation
field from the unshocked gas. Of course, this assumption breaks down close to the shock
front, \highlight{where fully three-dimensional calculations are needed}.
But it could also represent the case for which the shock loses all opacity
due to dust evaporation.

\begin{figure}
  \includegraphics[width=\linewidth]{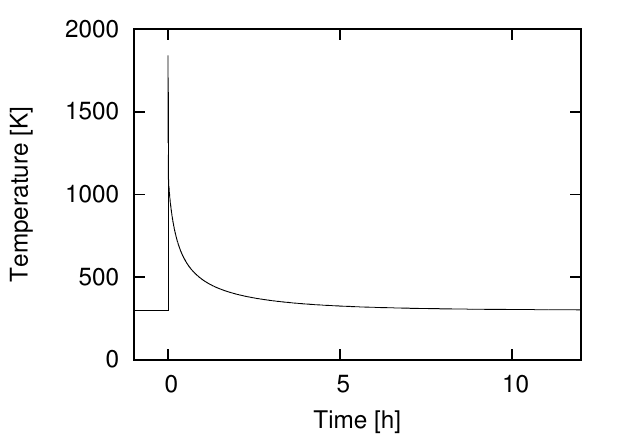}
  \caption{Particle temperature of an optically thin shock with  a shock
      speed of $V_\mathrm{shock}=9.0$ km/s. See text for details.}
  \label{gfxOpticallyThin}
\end{figure}

The result is shown in figure \ref{gfxOpticallyThin}. As expected there is no
pre-heating. Therefore the destination temperature for the jump conditions is
lower and therefore also the target temperature. To reach the melting
temperature in an optically thin case, we needed to increase the shock speed
to $V_\mathrm{shock}=9.0$ km/s, instead of $\sim 7.0$ km/s in the optically
thick cases.

Right after the shock the gas and particles cool down rapidly and approach
asymptotically the ambient temperature. The particles are only for a few
minutes at critical temperatures and are back below 500 K after $\sim 2$ h.

\begin{figure}
  \includegraphics[width=\linewidth]{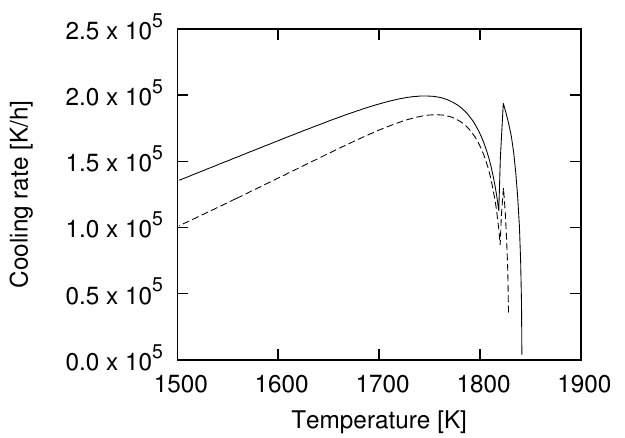}
  \caption{Cooling rates at $\mathrm{T} > 1500 ~\mathrm{K}$ in the optically
      thin simulation (solid line) and with the Planck-mean opacity decreased
      by a factor of 10$^{-3}$ (dashed line). The discontinuity at $\sim 1820 ~\mathrm{K}$ is
      due to the change of the heat capacity regime introduced earlier in section
      \ref{sectionResults}.}
  \label{gfxOpticallyThinCoolingRates}
\end{figure}

Unfortunately, the cooling rates (figure \ref{gfxOpticallyThinCoolingRates}, solid line)
are at least two orders of magnitude too high in the crystallization regime
at T $> 1500$ K to produce the observed chondrule textures \citep[see e.g.][]{hewins05,desch12}.
Since the Planck mean opacity is always an upper limit on the opacity we arbitrarily
decreased it by a factor of $10^{-3}$ to investigate the effect of lower opacities.
The gas has then a lower ability to directly cool by radiation. The major cooling
channel is then by thermal contact with the particles, which slows down the
overall cooling process. The result is shown in figure \ref{gfxOpticallyThinCoolingRates}
(dashed line). The cooling rates are still too large to be consistent with
chondrule formation. Decreasing the opacity even further does not have any effect, since
it is already low enough that the cooling via the particles is the dominant process.

In the bow shock scenario the optically thin approximation breaks down
close to the shock front. Therefore, in reality there will be a pre-heating just before
the shock front and less
cooling after the shock. Whether this can produce the desired cooling
rates has to be investigated by fully two-dimensional simulations. Further
-- but still one-dimensional -- simulations by \citet{morris12} suggest that this
could be the case.

\section{Conclusions}
\label{sectionConclusions}

Our 1-D radiative nebular shock model treats the variables at the downstream
boundary as output of the model instead of boundary conditions that can be
set. The iteration of the model then automatically finds the right
downstream state of the matter, given the upstream boundary conditions.  We
find that this procedure prevents a post-shock slow (few minutes) cooling
process. Instead, after the temperature spike and super-rapid cooling right
after the main shock, the temperatures stay virtually constant.  Only at
distances from the shock comparable to the scale height of the disk will the
1-D approximation break down and will sideways (i.e.\ upward and downward)
cooling set in. In our model we mimic this with a simple sideways cooling
term.

For cases where the shock structure is {\em local}, for instance the bow
shock of a planetesimal \citep[see][]{morris12}, the shock
scenario for chondrules might work because then the 1-D plane parallel
assumption breaks down and sideways cooling can commence in a matter of
hours. If the disk is optically thin then the cooling rates are too high to
be consistent with the constraints on chondrule formation.

We conclude that while the nebular shock model for chondrules may work for
local shocks (e.g. planetesimal bow shocks) the scenario has difficulties for
global shocks in an optically thick nebula. This is because such global radiative shocks do not produce
sufficient cooling after the temperature spike to be consistent with
meteoritic constrains.

\section{Acknowledgements}
\label{acknowledgements}

This work has been supported by the Deutsche Forschungsgemeinschaft
Schwerpunktprogramm (DFG SPP) 1385 ``The first ten million years of the solar
system''.

\highlight{We would like to thank Guy Libourel and an unknown referee for their
fruitful comments on this paper.}

\bibliographystyle{elsarticle-harv}
\bibliography{bibliography.bib}




%% \bibitem must have one of the following forms:
%%   \bibitem[Jones et al.(1990)]{key}...
%%   \bibitem[Jones et al.(1990)Jones, Baker, and Williams]{key}...
%%   \bibitem[Jones et al., 1990]{key}...
%%   \bibitem[\protect\citeauthoryear{Jones, Baker, and Williams}{Jones
%%       et al.}{1990}]{key}...
%%   \bibitem[\protect\citeauthoryear{Jones et al.}{1990}]{key}...
%%   \bibitem[\protect\astroncite{Jones et al.}{1990}]{key}...
%%   \bibitem[\protect\citename{Jones et al., }1990]{key}...
%%   \harvarditem[Jones et al.]{Jones, Baker, and Williams}{1990}{key}...
%%


% \end{thebibliography}

\appendix
\label{contentAppendix}

\section{Drag force/frictional heating}
\label{sectionAppendixDragForce}

If the particles are in a gas flow, which has a different velocity, they are
accelerated or decellerated, respectively, by exchanging momentum with the
gas. During this process they also get heated up by frictional heating of the
gas molecules, just like a spacecraft is heated up while re-entering the
Earth's atmosphere.
This was described by \citet[and references therein]{gombosi86}. If the
particle radius is smaller than the mean free path of the gas molecules --
which is valid up to a gas mass density of the order of $10^{-7}$ g/cm$^3$ for
millimeter-sized particles \citep{desch02} -- the drag coefficient
$C_\mathrm{D}$ used in equation (\ref{eqnDragForce}) is given by
\begin{equation}
  \begin{split}
    C_{\mathrm{D},j} = & \frac{2}{3s}\sqrt{\frac{\pi T_j}{T_\mathrm{g}}}
	+ \frac{2s^2+1}{\sqrt{\pi}s^3} \exp \left( -s^2 \right) \\
    & + \frac{4s^4+4s^2-1}{2s^4} \mathrm{erf} \left( s \right).
  \end{split}
\end{equation}
The parameter $s$ is the absolute value of the difference of the particle and
the gas velocity measured in units of sound speeds
\begin{equation}
  s = \frac{\left| V_j - V_\mathrm{g} \right|}
      {\sqrt{2k_\mathrm{B}T_\mathrm{g}/\overline{m}}},
\end{equation}
with the mean molecular weight $\overline{m}$. The \emph{error function} is
defined as
\begin{equation}
  \mathrm{erf}\left( x \right) = \frac{2}{\sqrt{\pi}}
      \int\limits_0^x e^{t^2}\mathrm{d}t.
\end{equation}

The recovery temperature used in equation (\ref{eqnFrictionalHeatingRate}) is
given by
\begin{equation}
  \begin{split}
    T_\mathrm{rec} = &T_\mathrm{g} \frac{\gamma - 1}{\gamma + 1} \left[
	\frac{2\gamma}{\gamma - 1} 2s^2 - \frac{1}{2} \right. \\
    + &\left. \frac{2}{\sqrt{\pi}} \exp \left( -s^2 \right) \mathrm{erf}^{-1}
	\left( s \right) \right]^{-1}.
  \end{split}
\end{equation}
If $s \rightarrow 0$, which means particles and gas share the same velocity,
then $T_\mathrm{rec} \rightarrow T_\mathrm{g}$, which means the particles
adapt the gas temperature with time. The heat transfer coefficient
$C_\mathrm{H}$ is given by
\begin{equation}
  \begin{split}
    C_{\mathrm{H},j} = & \frac{\gamma + 1}{\gamma - 1}
	\frac{k_\mathrm{B}}{8\overline{m}s^2} \times \\
    & \times \left[ \frac{s}{\sqrt{\pi}} \exp \left( -s^2 \right) + \left(
	\frac{1}{2} + s^2 \right) \mathrm{erf}\left( s \right) \right].
  \end{split}
\end{equation}

\section{Hydrodynamic calculations}
\label{sectionAppendixHydDynCalc}

Equations (\ref{eqnTotalMomentum}) and (\ref{eqnTotalEnergy}) have to be
extensively manipulated to get to the two coupled differential equations
(\ref{eqnCoupled1}) and (\ref{eqnCoupled2}) of the gas temperature and
velocity. Since this calculus is somewhat obscure we want to present it here
in all details.

Equation (\ref{eqnTotalMomentum}), which represents the momentum conservation
consists of the derivative of three terms. We want to do the derivative for
every term separately beginning with
\begin{equation}
  \begin{split}
    & \frac{\partial}{\partial x} \sum\limits_{i=1}^{N_I} \rho_i V_\mathrm{g}^2 =
	\frac{\partial}{\partial x} \sum\limits_{i=1}^{N_I} m_i n_i 
	V_\mathrm{g}^2 \\
    & \quad = \sum\limits_{i=1}^{N_I} m_i V_\mathrm{g} \frac{\partial}{\partial x}
	\left( n_i V_\mathrm{g} \right) + \sum\limits_{i=1}^{N_I} m_i n_i
	V_\mathrm{g} \frac{\partial}{\partial x} V_\mathrm{g}.
  \end{split}
  \label{eqnAppMomentum1}
\end{equation}
The first term here can be replaced using the continuity equation of the gas
(\ref{eqnGasContinuity}) while the second term is already in the needed form
only replacing the sum with the total gas mass density
\begin{equation}
  \begin{split}
    & \frac{\partial}{\partial x} \sum\limits_{i=1}^{N_I} \rho_i V_\mathrm{g}^2 =
	V_\mathrm{g} \sum\limits_{i=1}^{N_I} \sum\limits_{k=0}^{N_K} m_i R_{i,k} 
	+ \rho_\mathrm{g} V_\mathrm{g} \frac{\partial}{\partial x}
	V_\mathrm{g}.
  \end{split}
\end{equation}

Similar transformations can be applied to the second term of equation
(\ref{eqnTotalMomentum})
\begin{equation}
  \begin{split}
    & \frac{\partial}{\partial x} \sum\limits_{i=1}^{N_I} n_i k_\mathrm{B}
	T_\mathrm{g} \\
    & \quad = k_\mathrm{B} T_\mathrm{g} \sum\limits_{i=1}^{N_I}
	\frac{\partial}{\partial x} n_i + k_\mathrm{B} \sum\limits_{i=1}^{N_I}
	n_i \frac{\partial}{\partial x} T_\mathrm{g}.
  \end{split}
\end{equation}
As with equation (\ref{eqnAppMomentum1}) the continuity equation of the gas
(\ref{eqnGasContinuity}) can be used to manipulate the first term, while the
second is already in its final form
\begin{equation}
  \begin{split}
    & \frac{\partial}{\partial x} \sum\limits_{i=1}^{N_I} n_i k_\mathrm{B}
	T_\mathrm{g} \\
    \quad = &\frac{k_\mathrm{B} T_\mathrm{g}}{V_\mathrm{g}}
	\sum\limits_{i=1}^{N_I} \left( \sum\limits_{k=0}^{N_K} R_{i,k} - n_i
	\frac{\partial}{\partial x} V_\mathrm{g} \right) \\
    & + k_\mathrm{B} \sum\limits_{i=1}^{N_I} n_i \frac{\partial}{\partial x}
	T_\mathrm{g}.
  \end{split}
\end{equation}

The third term of equation (\ref{eqnTotalMomentum}) with the particle
momentum can be transformed as
\begin{equation}
  \begin{split}
    & \frac{\partial}{\partial x} \sum\limits_{j=1}^J n_j m_j V_j^2 \\
    & \quad = \sum\limits_{j=1}^{N_J} m_j V_j \frac{\partial}{\partial x} \left(
	n_j V_j \right) + \sum\limits_{j=1}^{N_J} n_j m_j V_j
	\frac{\partial}{\partial x} V_j \\
    & \quad + \sum\limits_{j=1}^{N_J} n_j V_j^2
	\frac{\partial}{\partial x} m_j.
  \end{split}
\end{equation}
The first term is equal to zero because of the continuity equation of the
particles (\ref{eqnParticlesContinuity}) and the second term can be replaced by
the drag force (\ref{eqnParticlesForce2}). The change in mass in the third
term can be transformed to a change in particle radius
\begin{equation}
  \begin{split}
    & \frac{\partial}{\partial x} \sum\limits_{j=1}^{N_J} n_j m_j V_j^2 \\
    & \quad = \sum\limits_{j=1}^{N_J} n_j F_{\mathrm{drag},j} +
	\sum\limits_{j=1}^{N_J} n_j V_j^2 4 \pi a_j^2 \rho_j
	\frac{\partial}{\partial x} a_j.
  \end{split}
\end{equation}
Adding all terms up this leads to equation (\ref{eqnCoupled1}).

Similar transformations can be applied to equation (\ref{eqnTotalEnergy}). We
do them here for all terms in the sums separately. The first term in the sum
over $i$ represents the kinetic energy of the gas due to its flow
\begin{equation}
  \begin{split}
    & \frac{\partial}{\partial x} \sum\limits_{i=1}^{N_I} \frac{1}{2} n_i m_i
	V_\mathrm{g}^3 \\
    & = \frac{1}{2} V_\mathrm{g}^2 \sum\limits_{i=1}^{N_I} m_i
	\frac{\partial}{\partial x} \left( n_i V_\mathrm{g} \right)
	+ V_\mathrm{g}^2 \sum\limits_{i=1}^I m_i n_i
	\frac{\partial}{\partial x} V_\mathrm{g}.
  \end{split}
\end{equation}
While the second term of this result is already in its final form, the first
term can be replaced by the continuity equation of the gas
(\ref{eqnGasContinuity})
\begin{equation}
  \begin{split}
    & \frac{\partial}{\partial x} \sum\limits_{i=1}^{N_I} \frac{1}{2} n_i m_i
	V_\mathrm{g}^3 \\
    & = \frac{1}{2} V_\mathrm{g}^2 \sum\limits_{i=1}^{N_I} \sum\limits_{k=0}^{N_K}
	m_i R_{i,k} + V_\mathrm{g}^2 \sum\limits_{i=1}^{N_I} m_i n_i
	\frac{\partial}{\partial x} V_\mathrm{g}.
  \end{split}
\end{equation}

The second term in the sum over $i$ in equation (\ref{eqnTotalEnergy}) is
the combined internal energy of the gas and the pressure. It can be
transformed as follows
\begin{equation}
  \begin{split}
    & \frac{\partial}{\partial x} \sum\limits_{i=1}^{N_I} n_i V_\mathrm{g}
	\frac{f_i + 2}{2} k_\mathrm{B} T_\mathrm{g} \\
    & = k_\mathrm{B} T_\mathrm{g} \sum\limits_{i=1}^{N_I} \frac{f_i + 2}{2}
	\frac{\partial}{\partial x} \left( n_i V_\mathrm{g} \right) \\
    & \quad + k_\mathrm{B} V_\mathrm{g} \sum\limits_{i=1}^{N_I} n_i \frac{f_i + 2}{2}
	\frac{\partial}{\partial x} T_\mathrm{g} \\
    & = k_\mathrm{B} T_\mathrm{g} \sum\limits_{i=1}^{N_I} \sum\limits_{i=0}^{N_K}
	\frac{f_i + 2}{2} R_{i,k} \\
    & \quad + k_\mathrm{B} V_\mathrm{g} \sum\limits_{i=1}^{N_I} n_i \frac{f_i + 2}{2}
	\frac{\partial}{\partial x} T_\mathrm{g},
  \end{split}
\end{equation}
where again the continuity equation of the gas (\ref{eqnGasContinuity}) was
used.

The first term in the sum over $j$ in equation (\ref{eqnTotalEnergy}) -- the
kinetic energy of the particles -- can be written as
\begin{equation}
  \begin{split}
    & \frac{1}{2} \frac{\partial}{\partial x} \sum\limits_{j=1}^{N_J} n_j m_j
	V_j^3 \\
    & = \frac{1}{2} \sum\limits_{j=1}^{N_J} m_j V_j^2 \frac{\partial}{\partial x}
	\left( n_j V_j \right) + \sum\limits_{j=1}^{N_J} n_j m_j V_j^2
	\frac{\partial}{\partial x} V_j \\
    & \quad + \frac{1}{2} \sum\limits_{j=1}^{N_J} n_j V_j^3
	\frac{\partial}{\partial x} m_j.
  \end{split}
\end{equation}
The first term here is again equal to zero because of equation
(\ref{eqnParticlesContinuity}). The second term can be replaced by the drag
force (\ref{eqnParticlesForce2}), while the third term can again be
transformed into a derivative of the particle radius
\begin{equation}
  \begin{split}
    & \frac{1}{2} \frac{\partial}{\partial x} \sum\limits_{j=1}^{N_J} n_j m_j
	V_j^3 \\
    & = \sum\limits_{j=1}^{N_J} n_j V_j F_{\mathrm{drag},j} + 2 \pi 
	\sum\limits_{j=1}^{N_J} n_j V_j^3 a_j^2 \rho_j
	\frac{\partial}{\partial x} a_j.
  \end{split}
\end{equation}

The second term in the sum over $j$ in equation (\ref{eqnTotalEnergy}) is the
internal energy of the particles
\begin{equation}
  \begin{split}
    & \frac{\partial}{\partial x} \sum\limits_{j=1}^{N_J} n_j V_j m_j
	C_{\mathrm{P},j} T_j \\
    & = \sum\limits_{j=1}^{N_J} m_j C_{\mathrm{P},j} T_j
	\frac{\partial}{\partial x} \left( n_j V_j \right) \\
    & \quad + \sum\limits_{j=1}^{N_J} n_j V_j \frac{\partial}{\partial x} \left(
	m_j C_{\mathrm{P},j} T_j \right) \\
    & = \sum\limits_{j=1}^{N_J} n_j V_j \frac{\partial}{\partial x} \left( m_j
	C_{\mathrm{P},j} T_j \right),
  \end{split}
\end{equation}
where again the continuity equation of the particles
(\ref{eqnParticlesContinuity}) was used. The remaining term can be further
transformed to
\begin{equation}
  \begin{split}
    & \frac{\partial}{\partial x} \sum\limits_{j=1}^{N_J} n_j V_j m_j
	C_{\mathrm{P},j} T_j \\
    & = \sum\limits_{j=1}^{N_J} n_j V_j m_j C_{\mathrm{P},j}
	\frac{\partial}{\partial x} T_j \\
    & \quad + 4 \pi \sum\limits_{j=1}^{N_J} n_j V_j C_{\mathrm{P},j} T_j a_j^2
	\rho_j \frac{\partial}{\partial x} a_j.
  \end{split}
\end{equation}

The derivative of the radiative flux $F_\mathrm{rad}$ is already given in
equation (\ref{eqnFrad}). Summing up every term, this yields equation
(\ref{eqnCoupled2}). The formulas for $\frac{\partial}{\partial x} T_j$ and
$\frac{\partial}{\partial x} a_j$ can be inserted from equations
(\ref{eqnParticlesTemperature}) and (\ref{eqnParticlesRadius}). Equations
(\ref{eqnCoupled1}) and (\ref{eqnCoupled2}) are two coupled differential
equations of the form
\begin{equation*}
  \begin{split}
    A\frac{\partial}{\partial x} V_\mathrm{g} + B\frac{\partial}{\partial x}
	T_\mathrm{g} &= C \\
    D\frac{\partial}{\partial x} V_\mathrm{g} + E\frac{\partial}{\partial x}
	T_\mathrm{g} &= F,
  \end{split}
\end{equation*}
with the respective $A,\ B,\ C,\ D,\ E$ and $F$. Decoupled they can be written
as
\begin{equation*}
  \begin{split}
    \frac{\partial}{\partial x} V_\mathrm{g} &= \frac{CE - BF}{AE - BD} \\
    \frac{\partial}{\partial x} T_\mathrm{g} &= \frac{AF - CD}{AE - BD}.
  \end{split}
\end{equation*}

\end{document}